\documentclass[a4paper]{article}

\usepackage{INTERSPEECH2020}
\usepackage{hyperref}

\title{ConVoice: Real-Time Zero-Shot Voice Style Transfer \\ with Convolutional Network}
\name{Yurii Rebryk, Stanislav Beliaev
\thanks{Preprint. Submitted to INTERSPEECH.}
}
\address{
  Higher School of Economics, Saint Petersburg, Russia
}
\email{y.a.rebryk@gmail.com, stasbelyaev96@gmail.com}

\begin{document}

\maketitle

\begin{abstract}
  We propose a neural network for zero-shot voice conversion (VC) without any parallel or transcribed data. Our approach uses pre-trained models for automatic speech recognition (ASR) and speaker embedding, obtained from a speaker verification task. Our model is fully convolutional and non-autoregressive except for a small pre-trained recurrent neural network for speaker encoding. ConVoice can convert speech of any length without compromising quality due to its convolutional architecture.  Our model has comparable quality to similar state-of-the-art models while being extremely fast. 
\end{abstract}

\noindent\textbf{Index Terms}: voice style transfer, voice conversion, convolutional networks


\section{Introduction}

Voice style transfer or voice conversion (VC) is a technique to make the speech of one speaker sound like a different speaker. Voice conversion has a wide range of practical applications such as privacy protection, content generation for the entertainment industry, and personalized speech synthesis.

In general, there are four major factors that define human speech: timbre, content, pitch and rhythm~\cite{speech_decomposition}. Timbre defines the speaker's voice characteristics, content carries the information in the speech, and pitch and rhythm describe its prosody. 
The majority of VC methods focus on timbre conversion, transforming the source utterance frame-by-frame, while keeping the prosody from the original speech\cite{ttsskins, starganvc, cyclegan}. In this case, the source and synthesized speech have the same duration. There are also models which map a source speech to a variable length output by using recurrent neural networks (RNNs) or attention-based auto-regressive models\cite{vc_s2s, vc_transformer, convs2svc}. Although the last methods are able to keep the prosody of the target speaker voice, they are usually slower and perform poorly on long speech samples.

There are two primary approaches to the voice style transfer problem: parallel and non-parallel VC. The parallel VC approach leverages a parallel speech dataset consisting of utterance pairs from source and target speakers. In practice, such a dataset is difficult, time-consuming and expensive to build.


The non-parallel VC approach does not require a parallel corpus. Many non-parallel VC models were inspired by image style transfer ideas in computer vision, such as Generative Adversarial Networks (GANs)\cite{starganvc, cyclegan, voicegan} and Variational Autoencoders (VAEs)\cite{cyclic_vae, oneshot_vae}. However, GANs are difficult to train, while VAEs' outputs tend to be over-smoothed\cite{starganvc} and do not produce high-quality human-like speech.
Another non-parallel VC approach is based on the encoder-decoder paradigm. The encoder uses an automatic speech recognition (ASR) model to extract linguistic information from the source utterance, and the decoder synthesizes new speech using the target speaker's voice\cite{ttsskins}. The downside of this approach is that a pre-trained ASR model is required. Recently, conditional auto-encoder (CAE) based methods were applied to this task\cite{autovc, f0} which capitalizes on a carefully crafted bottleneck that separates the speaker's identity from the source utterance, and uses a speaker embedding as an input to the decoder model.

Also, it is worth mentioning that some approaches described above can do voice conversion only with a speaker from the training dataset. Other methods can be fine-tuned with a relatively small amount of data. There are several methods which perform zero-shot conversion, i.e. conversion to the voice of a new speaker using only a handful of target utterances.

Our method is closely related to the TTS Skins\cite{ttsskins} model. TTS Skins is an encoder-decoder network which uses acoustic features extracted from the pre-trained ASR model Wav2Letter\cite{openseq2seq}. The main differences between TTS Skins and our model are: (1) we use a modern QuartzNet-5x5\cite{kriman2019quartznet} encoder that is 20 times smaller that Wav2Letter, (2) we use a separate fully-convolutional decoder and pre-trained non-autoregressive WaveGlow\cite{waveglow} vocoder instead of WaveNet, and (3) we use a speaker encoder that is pre-trained on a speaker verification task to extract speaker embeddings\cite{jia2018transfer} instead of using a lookup table, making zero-shot conversion possible. 

Thus, our model is significantly smaller, more computationally efficient and does not require the collection of new data when converting to a new speaker's voice.

In this paper, we present an encoder-decoder network that does frame-by-frame zero-shot voice conversion without parallel or transcribed data.
Our method utilizes pre-trained ASR and speaker encoder models. Our model is fully convolutional and non-autoregressive except for the small pre-trained RNN for the speaker encoder. Moreover, the model can convert sound of any length without compromising quality and perform inference much faster than real-time.

\begin{figure}[t]
  \centering
  \includegraphics[width=\linewidth]{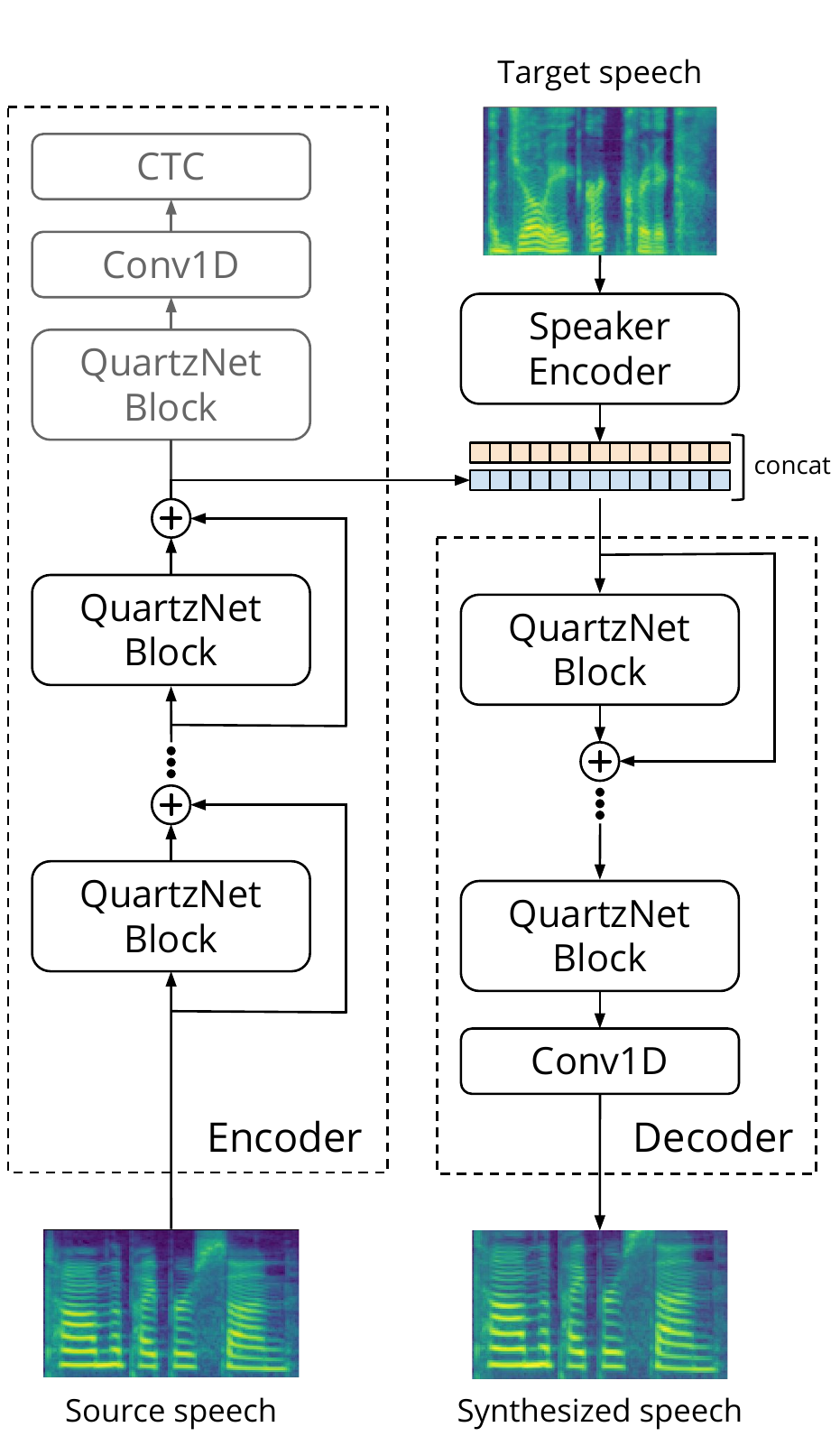}
  \caption{\textbf{The architecture of ConVoice model.} The pre-trained ASR model extracts acoustic features from the source utterance and the pre-trained speaker encoder produce the target speaker embedding from the target audio. After these embeddings are concatenated and fed into the decoder that generates the mel-scale log spectrogram.}
  \label{fig:convoice_model}
\end{figure}

\section{Model architecture}

Our model is based on an encoder-decoder paradigm and consists of four independently trained neural networks (see Fig.~\ref{fig:convoice_model}):
\begin{itemize}
    \item \textbf{Audio encoder}: A convolutional network based on QuartzNet~\cite{kriman2019quartznet} which extracts audio features from speech input.
    \item \textbf{Speaker encoder}: An RNN based on \cite{wan2017generalized} which computes a fixed dimensional vector that represents a speaker embedding.
    \item \textbf{Decoder}: A convolutional decoder which predicts a mel-spectrogram from encoder acoustic features conditioned on the speaker embedding.
    \item \textbf{Vocoder}: A WaveGlow vocoder which synthesizes speech from the mel-spectrogram.
\end{itemize}

\subsection{Audio encoder}
The purpose of an audio encoder is to transform speech into a representation suitable for the synthesis of the original speech in the voice of another speaker, if being conditioned on the speaker embedding. It is crucial to use a representation which captures enough information about sounds pronounced in the original speech but does not contain too much details about the characteristics of the original speaker's voice. \\

We use QuartzNet-5x5 model for ASR. The model maps a mel-spectrogram into a sequence of symbols. QuartzNet-5x5 is represented by 1D convolutional layer followed by five basic blocks with residual connections between them. Each block consists of the same module that is repeated five times and contains the following layers: 1) a 1D time-channel separable convolutional layer, 2) a batch normalization layer, and 3) ReLU. This model is trained using Connectionist Temporal Classification (CTC) loss\cite{10.1145/1143844.1143891} and achieves near state-of-the-art accuracy on LibriSpeech and Wall Street Journal datasets. 

Despite the fact that the model was trained for speech recognition task, we find that, similarly to \cite{ttsskins}, the embeddings from intermediate layers can be used by the decoder to synthesize the original speech in another voice. We use embeddings that are produced by the fourth QuartzNet block.

The input of the encoder is a 80-channels mel-scale log-magnitude spectrogram with a 1024 window size and a 256 hop which is normalized separately per each mel channel.

\subsection{Speaker encoder}

The speaker encoder should be able to map a short speech utterance to a vector of fixed dimension which captures the characteristics of a speaker voice. As shown in \cite{jia2018transfer}, a speaker-discriminative model trained on a text-independent speaker verification task can produce an embedding that is suitable for conditioning the decoder on speaker identity.

The speaker encoder consists of a 3-layer LSTM of 256 cells followed by a fully connected layer of 256 units, as suggested in \cite{wan2017generalized}.

The model is trained using a generalized end-to-end speaker verification loss, that forces the network to produce embeddings with high cosine similarity if utterances belong to the same speaker, and low cosine similarity if they do not.

The input to the network is a 40-channel mel-scale log spectrogram of 1.6 seconds of utterance with a 25ms window size and a 10ms hop. The output is $L_2$-normalized output of the top LSTM layer at the final frame.

During inference, utterances are split into segments of 1.6 seconds overlapping by 50\% which are fed into the encoder separately. These embeddings are averaged and normalized to form the final speaker embedding.

\subsection{Decoder}
The decoder is used to generate mel-spectrogram from encoder embeddings, conditioned on a speaker embedding. 

Our decoder is a smaller version of QuartzNet. Our model has only three basic blocks with residual connections between them and uses kernels of smaller width, which is roughly 1/4 of the original QuartzNet size. The basic block architecture consisting of 1D time-channel separable convolution, batch normalization and ReLU, is unchanged.

The decoder is trained using a pre-trained audio encoder and a pre-trained speaker encoder to extract audio embeddings and the target speaker embedding, respectively. Parameters of the audio encoder and the speaker encoder are frozen during the training. The audio embeddings are concatenated with the target speaker embedding at each time step and fed into the decoder network. 

The target is an 80-channel mel-scale log magnitude spectrogram with a 1024 window size and a 256 hop that is computed from a 22,050Hz audio signal. We train the model using $L_2$ loss.

\subsection{Vocoder}
We use WaveGlow as a vocoder to synthesize waveforms from mel-spectrograms, because it is fast and provides high-quality audio.

The input of the network is a 80-channel mel-spectrogram with a 1024 window size and a 256 hop. The output of the vocoder is an audio signal with 22,050Hz sampling rate.

\section{Experiments}
The audio encoder was pre-trained using LibriTTS\cite{libritts} dataset, which consists of approximately of $585$ hours of read English speech. Selected configuration of the encoder performs worse than original QuartzNet-15x5\cite{kriman2019quartznet}, but it has almost $3$ times fewer parameters. The model achieves word-error-rate (WER) of 8.36\% on LibriTTS test-clean and 18.40\% LibriTTS test-other sets.

We used a pre-trained speaker encoder from an open-source GitHub repository\footnote{\url{https://github.com/CorentinJ/Real-Time-Voice-Cloning}} that was trained on VoxCeleb1\cite{voxceleb1}, VoxCeleb2\cite{voxceleb2}, and LibriSpeech\cite{librispeech} datasets.
VoxCeleb1 and VoxCeleb2 datasets consist of utterances extracted from YouTube videos of celebrities. Both datasets have about $7,200$ speakers, but their audio segments are very noisy.
LibriSpeech dataset contains about $1,000$ hours of audio that are nearly evenly split into ``clean'' and ``other'' sets. Only the ``other'' set was used for training. The authors filtered non-English speakers from VoxCeleb1, and downsampled all datasets at 16kHz.

The decoder network was trained using LibriTTS dataset that was downsampled to 22,050Hz. The train set of the dataset is divided into ``clean'' and ``other'' sets, that have $1,151$ and $1,160$ speakers, and contain $245$ and $310$ hours of speech respectively. But we use only ``clean'' set for training.

Admittedly, as LibriTTS dataset was used for the training of both the audio encoder and decoder models, this approach can be harmful to the audio quality of synthesized speech because the distribution of audio embeddings that were seen during the training could differ from those, that are used during the inference.
However, it was done due to a lack of publicly available large audio datasets containing high-quality speech produced by a diverse set of speakers.


To convert predicted mel-scale log-magnitude spectrograms to waveforms, we used WaveGlow vocoder that was pre-trained on the LJ Speech dataset\cite{ljspeech}.

We evaluated the proposed method on the Voice Conversion Challenge 2018 (VCC2018)\cite{vcc2018}. This challenge consists of Hub and Spoke tasks, in which participants should synthesize source utterance with the target voices. Both tasks have four different source speakers (2 male and 2 female) and four common target speakers (2 male and 2 female). The training dataset contains about five minutes of speech for each speaker. The Hub and Spoke tasks are similar, except that the training data for the Hub task is parallel corpora. The test set of each task consists of $35$ utterances for each source speaker that should be converted to each target voice. Thus, there are $35 \times 4 \times 4 = 560$ audio samples that should be synthesized per each task.
We conducted the Mean Opinion Score (MOS) evaluation on Amazon Mechanical Turk to measure the naturalness and speaker similarity of synthesized speech, and compared our method with N10\cite{n10}, which is the best system of VCC2018. Our samples are available on the demo page\footnote{\url{https://rebryk.github.io/convoice-demo/}}.

\begin{table}[th]
  \caption{Speech naturalness MOS with 95\% confidence intervals.}
  \label{tab:naturalness}
  \centering
  \begin{tabular}{lrr}
    \toprule
    \multicolumn{1}{c}{\textbf{Method}} & \multicolumn{1}{c}{\textbf{Hub}} & \multicolumn{1}{c}{\textbf{Spoke}} \\
    \midrule
    GT source               & $4.41 \pm 0.11$ & $4.34 \pm 0.11$ \\
    GT target               & $4.49 \pm 0.11$ & $4.49 \pm 0.11$ \\
    N10                     & $3.75 \pm 0.08$ & $3.90 \pm 0.08$ \\
    \midrule
    ConVoice (zero-shot)    & $3.72 \pm 0.09$ & $3.72 \pm 0.08$ \\
    ConVoice                & $3.94 \pm 0.13$ & $3.92 \pm 0.12$ \\
    \bottomrule
  \end{tabular}
\end{table}

\subsection{Speech naturalness}
We synthesized two sets of samples. The first set was produced in zero-shot setting, when the model has not seen the target or source speaker before, and the second one is created using the model fine-tuned on the VCC2018 training dataset. To measure the speech naturalness of the synthesized audio, each utterance was evaluated by one person, who rated its naturalness on a five-point scale. The results are shown in Table~\ref{tab:naturalness} with 95\% confidence interval. 


As you can see, in zero-shot setting ConVoice performs slightly worse than N10, because it generates some background noise. However, after fine-tuning of the model on a small amount of data containing utterances pronounced by the target speakers, this unpleasant noise nearly disappears and synthesized speech quality improves, matching the performance of N10.

\subsection{Speaker similarity}
We compared speaker similarity of the generated and ground truth audio similar to VCC2018. We asked raters to indicate how sure they are that the given samples are produced by the same speaker, despite some optional distortion of the audio. Raters had four options to choose from that are presented in Table~\ref{tab:ratings}. The Mean Opinion Scores are shown in Table~\ref{tab:similarity}.

\begin{table}[th]
  \caption{Ratings that were used in evaluation of similarity of synthesized and ground truth samples.}
  \label{tab:ratings}
  \centering
  \begin{tabular}{ccc}
    \toprule
    \multicolumn{1}{c}{\textbf{Rating}} & \multicolumn{1}{c}{\textbf{Similarity}} & \multicolumn{1}{c}{\textbf{Confidence}} \\
    \midrule
    1 & Not same    & Absolutely sure \\
    2 & Not same    & Not sure \\
    3 & Same        & Not sure \\
    4 & Same        & Absolutely sure \\
    \bottomrule
  \end{tabular}
\end{table}

The speech similarity of samples that were generated by the ConVoice model in zero-shot setting is also not so high in comparison with the fine-tuned model. Two possible explanations of this result are: 1) probably, it is more difficult to compare voices when audio contains noise, that we mentioned above, and 2) obviously, the fine-tuned model should know better how to synthesize a voice of a speaker, that was included in the training data. Speech similarity MOS for the fine-tuned ConVoice and N10 are almost the same.

\begin{table}[th]
  \caption{Speech similarity MOS with 95\% confidence intervals.}
  \label{tab:similarity}
  \centering
  \begin{tabular}{lcc}
    \toprule
    \multicolumn{1}{c}{\textbf{Method}} & \multicolumn{1}{c}{\textbf{Hub}} & \multicolumn{1}{c}{\textbf{Spoke}} \\
    \midrule
    GT source               & $1.30 \pm 0.06$ & $1.24 \pm 0.06$ \\
    GT target               & $3.96 \pm 0.05$ & $3.96 \pm 0.05$ \\
    N10                     & $3.30 \pm 0.09$ & $3.31 \pm 0.09$ \\
    \midrule
    ConVoice (zero-shot)    & $2.93 \pm 0.12$ & $2.88 \pm 0.09$ \\
    ConVoice                & $3.30 \pm 0.09$ & $3.39 \pm 0.08$ \\
    \bottomrule
  \end{tabular}
\end{table}

Also, we measured how speaker's gender affects the speech similarity of the generated audio (Table~\ref{tab:similarity_gender}). From these results, we cannot conclude that the gender of the speaker significantly affects synthesized speech. We can see that the voice conversion to a male voice has slightly better results. However, the confidence intervals are too wide.

\begin{table}[th]
  \caption{Speech similarity MOS (with 95\% confidence intervals) of the synthesized and ground truth audio based on the gender of the speakers. Samples were produced using the fine-tuned ConVoice model.}
  \label{tab:similarity_gender}
  \centering
  \begin{tabular}{llcc}
    \toprule
    \multicolumn{1}{c}{\textbf{Source}} & \multicolumn{1}{c}{\textbf{Target}} & \multicolumn{1}{c}{\textbf{Hub}} & \multicolumn{1}{c}{\textbf{Spoke}} \\
    \midrule
    Male    & Male      & $3.30 \pm 0.24$ & $3.24 \pm 0.27$ \\
    Male    & Female    & $3.22 \pm 0.25$ & $3.14 \pm 0.23$ \\
    Female  & Male      & $3.42 \pm 0.13$ & $3.49 \pm 0.13$ \\
    Female  & Female    & $3.19 \pm 0.16$ & $3.43 \pm 0.14$ \\
    \bottomrule
  \end{tabular}
\end{table}

\subsection{Inference speed}

\begin{table}[th]
  \caption{Inference latency of ConVoice model averaged over VCC2018 dataset. Single V100 GPU. Average sample length is about 3.5s. Number of tested samples is 512. Latency and real-time factor are calculated without vocoding.}
  \label{tab:latency}
  \centering
  \begin{tabular}{ccc}
    \toprule
    \multicolumn{1}{c}{\textbf{Batch size}} & \multicolumn{1}{c}{\textbf{Inference latency, s}} & \multicolumn{1}{c}{\textbf{RTF}} \\
    \midrule
    1      & $0.011 \pm 8 \cdot 10^{-4}$    & $321.58$ \\
    4      & $0.013 \pm 16 \cdot 10^{-4}$   & $1077.79$ \\
    8      & $0.020 \pm 37 \cdot 10^{-4}$   & $1422.49$ \\
    \bottomrule
  \end{tabular}
\end{table}

Our encoder and decoder networks are lightweight and have 11M parameters in total. As you can see in Tab.~\ref{tab:latency} our model without the vocoder is extremely fast. However, if one wants to synthesize speech using WaveGlow vocoder, the inference speed is drastically lower. WaveGlow model has 87.7M parameters, and is the bottleneck of our method. Despite that, the whole pipeline is still faster than real-time and produces about 12 seconds of synthesized 20,050Hz audio in 1 second of wall time on an NVIDIA V100 GPU. In a future work WaveGlow can be replaced with a smaller and more efficient model such as WaveFlow\cite{waveflow} or SqueezeWave\cite{squeezewave}.

\section{Conclusions}
We proposed a encoder-decoder NN which produces zero-shot voice conversion without parallel and transcribed data, using pre-trained ASR and speaker encoder models. 

Our model is compact, having only 11M parameters without the vocoder. Convolutional architecture of the network makes it easy to parallelize and helps achieve much faster inference speed. 

Besides that, the model converts speech of any length without compromising quality. We conducted the MOS evaluation to measure naturalness and speaker similarity of synthesized speech, and our method showed decent results.

The ability of our approach to perform zero-shot voice conversion, given a few seconds of the target speaker voice, can be crucial in applications, where customers deploy the model on their side and prefer not to share private voices they use, or if collecting of a new dataset is sophisticated. 
\section{Acknowledgments}

The authors thank Boris Ginsburg and Oleksii Hrinchuk for their helpfull discussions and advice, Roman Kotenko for the feedback and review, as well as Sergey Golovanov for providing credits for Amazon Mechanical Turk.

\bibliographystyle{IEEEtran}
\bibliography{mybib}

\end{document}